# Super-resolution imaging and microscopy by dielectric particle-lenses

**Zengbo Wang[1], Boris Luk'yanchuk[2,3]**

[1] School of Computer Science and Electronic Engineering, Bangor University, Bangor LL57 1UT, UK

[2] Division of Physics and Applied Physics, School of Physical and Mathematical Sciences, Nanyang Technological University, Singapore 637371, Singapore

[3] Faculty of Physics, Lomonosov Moscow State University, Moscow 119991, Russia

**Abstract**  Recently, imaging by microspheres and dielectric particle-lenses emerged as a simple solution to obtaining super-resolution images of nanoscale devices and structures. Calibrated resolution of ~λ/6 - λ/8 has been demonstrated, making it possible to directly visualize 15-50 nm scale objects under a white light illumination. The technique has undergone rapid developments in recent years, and major advances such as the development of surface scanning functionalities, higher resolution metamaterial superlens, biological superlens and integrated bio-chips as well as new applications in interferometry, endoscopy and others, have been reported. This paper aims to provide an overall review of the technique including its background, fundamentals and key progresses. The outlook of the technique is finally discussed.

## 1 Introduction

### 1.1 Background

Optical microscopy is one of the most important scientific achievements in the history of mankind. The invention of the compound optical microscope by Hans & Zaccharis Janssen in 1590 and improvements by Galileo Galilei, Robert hook, Anthony Leeuwenhoek has revolutionized all aspects of science and technology, especially the life sciences when it became possible for researcher to see, for instance, the bacteria and blood cells.

In 1873, Ernst Abbe established the resolution limit of optical microscopes[1]: The minimum distance, d', between two structural elements to be imaged as two objects instead of one is given by d' = Kλ/NA=0.5λ/NA, where λ is the wavelength of light and NA the numerical aperture of the objective lens. Other ways of resolution definition is possible, such as Sparrow[2], Houston[3] or Rayleigh[4] criteria, where K=0.473,0.515, or 0.61 respectively. The physical root of resolution limit is related to optical diffraction and loss of evanescent waves in far-field; the evanescent waves carry high-frequency subwavelength spatial information of an object and decay exponentially with distance from the object. With white lights, optical microscope resolution is limited at about 200-250 nm.  For about one hundred years, the resolution criterion was considered the fundamental limit of optical microscope resolution.

One key step forward in beating diffraction limit is the invention of near-field scanning optical microscopy (NSOM) technique by D.W. Pohl and colleagues in 1984. The NSOM allows sub-wavelength optical imaging for the first time[5]. An image of a structure is constructed by scanning a physical tip with subwavelength aperture in the proximity (~tens of nanometres) of an illuminated specimen. Since the late 1990s, stimulated by the surge of nanophotonics, plasmonics and metamaterials, a number of new super-resolution microscopy/nanoscopy techniques have appeared which include metamaterial superlens[6] and STED (stimulated emission depletion microscopy)[7].

Metamaterial superlens, or Pendry superlens, was first theoretically proposed by British scientist John Pendry in 2000 which use a slab of NIM (negative-index medium) to enhance the evanescent waves, offers the possibility to restore the nanoscale information in the far-field and therefore a nearly perfect image can be reconstructed[6]. The 'perfect lens' concept was widely taken by the community and sparked a research surge in metamaterials and plasmonics. Several versions of plasmonic metamaterial superlenses, which follow Pendry's basic idea, have been developed and demonstrated by several groups and researchers across the world in the past decade[8-11]. It is important to note that in its currently most advanced version, called hyper-lens[9,10], the evanescent waves are converted into propagating waves forming a magnified image of the sample on a distant screen, which is why one may think that it is far-field imaging. However, the projection to a distant screen does not change the fact that the hyper lens relies on the sample's near-field. Hence, in its current state of development, a hyper-lens is not a far-field imaging device[9], but a non-scanning concept of employing the near-field. Because of the fundamental loss limit of plasmonic materials used in these lenses, and nanofabrication challenges, resolutions are still limited at 70~100 nm at a single visible wavelength after a decade's efforts by researchers across the world and haven't been well appreciated for bio-imaging application.

Into the far-field domain, the super-resolution fluorescence microscopy techniques are possibly the most successful developments in recent year. They are attained with two main approaches: spatially patterned excitation (STED, RESOLFTs, SSIM)[12] and single-molecule localization (STORM, PALM, FPALM)[12] of fluorescence molecules. The techniques are already applied on a large scale in major fields of the biological sciences, like cell biology, microbiology and neurobiology, and may revolutionize the entire biology and medicine fields in the future. The developers of these techniques (Erik Betzig, Stefan W. Hell and W. E. Moerner) were awarded the Nobel Prize in Chemistry in 2014. But since these techniques are fluorescence based, they are not applicable for imaging non-fluorescence samples, including for example the semiconductor chip devices and biological viruses and sub-cellular structures which cannot be labelled using existing fluorophores. Besides, these techniques are not based on high-resolution lens, but on fluorescent materials. The use of fluorescence may also change original function of researched biological objects and affect its biological dynamic processes as well. Therefore, there still exists the strong need to develop a super-resolution lens which can offer label-free, high-resolution imaging of different samples.

Recently, through manipulation of the diffraction of light with binary masks or gradient met surfaces, some miniaturized and planar lenses have been reported with intriguing functionalities such as subdiffraction-limit focusing in far-field, ultrahigh numerical aperture and large depth of focus, which provides a viable solution for the label-free super-resolution imaging. The most well-known example is possibly the 'superoscillatory lens (SOL)' by Zheludev's group in Southampton[13]. The basic which has root connection with Toraldo di Francia's proposal in 1956[14], is to use a carefully-designed amplitude or phase zone plate to modulate the beam to achieve a super-resolution spot in the far-field, through constructively and destructive interference without evanescence waves being involved. The key disadvantage in the technique is the appearance of giant sidelobes near to the central spot, which affected the practical adoption of the technology. The central spot resolution can, in theory, range from infinite small to $0.38\lambda/NA$ (known as 'super-oscillation criteria' in [15]; NA: Numerical Aperture). More recently, Qin et al. reported the development of supercritical lens (SCL), a planar diffraction component which has a focusing spot smaller than $0.61\lambda/NA$ but slightly larger than SOL ($0.38\lambda/NA$), and a needle-like focal region with its Depth of Focus (DOF) $z = 2\lambda/NA^2$ that differs from traditional spherical lens, Fresnel Zone Plate, SOL, and others [16]. The main advantage of SCL is its 3D imaging capability (DOF: $12\lambda$) in axial direction with modest super-resolution ($0.41\lambda$) in lateral direction. For more information on latest development of planar diffractive lens, please refer to a latest review by Huang and co-workers[15]. In addition, several other super-resolution techniques and proposals exist in the literature, including structured illumination, Maxwell fisheyes, scattering lens and time-reversal imaging. More details regarding these technique can be found in an magazine article written by Cartwright[17].

## 1.2 Microsphere super-resolution imaging

Wang and co-workers published their pioneering work on microsphere nanoscopy in 2011 [18]. The technique uses micro-sized spheres as super-resolution lenses (superlens) to magnify underlying objects before projecting them into a conventional microscope's objective lens (see Fig. 1). The spheres generate sub-diffraction illumination on the underlying object, excite and collect the near-field evanescent object information and form virtual images that are subsequently captured by the conventional lens. This is a label-free and real-time imaging technique, which can directly resolve ~50 nm features under white light illumination. Such resolving power corresponds to a calibrated resolution of ~ $\lambda/6$ - $\lambda/8$ based on convolution fitting [19]. These features are unique and attractive for achieving low-intensity high-resolution imaging of any objects[20-22,18,23,24] in nanometre sizes and has been successfully demonstrated for both biological (cells, viruses, etc. [22] [25]) and non-biological samples (semiconductor chips, nanoparticles etc.).

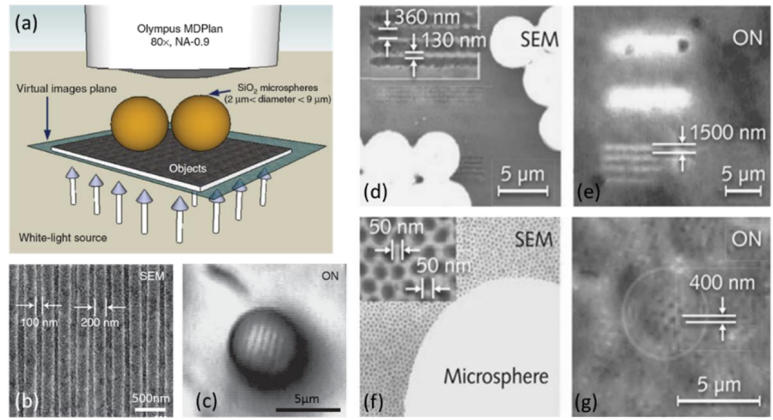

Fig. 1 (a)Experimental configuration of microsphere superlens integrated with a classical optical microscope. (b-g) Experimental results for microsphere superlens imaging. (b)SEM image of a Blu-ray DVD disc (200 nm lines and 100 nm grooves). (c) Optical imaging by microsphere in reflection mode. (d) shows SEM image of a diffraction grating with 300 nm wide lines spaced 130 nm apart. (e) shows the lines are clear resolved by microsphere with a 4.17X magnification in transmission mode. (f) shows the SEM image of a gold-coated fishnet membrane sample with 50 nm pores spaced 50 nm apart. (g) shows the optical image by microsphere superlens with 8X magnification in transmission mode[14].

Figure 1(a) shows the typical experimental setup of *'microsphere nanoscope'* in transmission mode. The $SiO_2$ microspheres, with diameter between 2 and 9 μm, are placed on top of the object's surface through self-assembly process. The as-received $SiO_2$ microsphere suspension is diluted and applied to the imaging samples by drop or dip coating and the samples are left to dry in air. A halogen lamp with a peak wavelength of 600 nm is used as the white-light illumination source. These microspheres function as superlenses - *microsphere super lenses (MSL)*-that collect the underlying near-field object information and magnify it (forming virtual images that keep the same orientation as the objects in the far-field) before it is projected to an 80X Olympus objective lens (NA = 0.9, model MDPlan) of an Olympus microscope (model MX-850). The combination of microsphere-superlenses (MSL) and the objective lens forms a compound-imaging lens system. In the reflection mode, the white-light source will be incident from the top, opposite to the light source at the bottom in the transmission mode.

Clear images of subdiffraction nanoscale objects have been captured by MSL in either transmission or reflection mode. Fig. 1(b and c) reveals that a Blu-ray DVD disc with 200-nm-wide lines and 100-nm-wide grooves are clearly imaged by a 4.7 μm microsphere in reflection mode. In another sample (Fig. 1 d and e), 30-nm thick chrome-film diffraction gratings with 360-nm-wide lines spaced 130 nm apart on fused silica substrates were imaged in transmission mode. The virtual image plane was 2.5 μm beneath the substrate surface, and only those lines with microsphere particles on top of them were resolved. The lines without the particles mix together and form a bright spot that cannot be directly resolved by the optical microscope because of the diffraction limit. For the visible wavelength 400 nm, the best diffraction-limited resolution is es-timated to be 215 nm in air using the vector theory of Richards and Wolf [26], and 152 nm when taking the solid-immersion effect of a particle into account. For the main peak of a white-light source at 600 nm, the limits are 333 nm in air and 228 nm with solid-immersion effect, respectively. Here, one should also note that the focal planes for the lines with and without particles on top are different.

In another example (Fig. 1 f and g), a fishnet 20-nm-thick gold-coated anodic aluminium oxide (AAO) membrane fabricated by two-step anodizing in oxalic acid (0.3 mol/l) under a constant voltage of 40 V is imaged with 4.7-μm-diameter microspheres. The membrane pores are 50 nm in diameter and spaced 50 nm apart. The microsphere nanoscope resolves these tiny pores well beyond the diffraction limit with a resolution about $\lambda/8$ in the visible. It is important to note that the magnification in this case is around 8X—almost two times that of the earlier grating example, implying that the performance of the microsphere super lens is affected by the near-field interaction of the sphere and the substrate. The sample-dependent resolution brings in complexity in clarifying the physical mechanism of the technique, but the overall resolution range is between 50 nm to 100 nm, well exceeding the classical limit of ~200 nm.

### 1.3 From microsphere to metamaterial particle-lens

The simplicity and high resolution of the microsphere imaging technique have attracted considerable interests. The technique was validated and resolution levels of ~ $\lambda/6$ - $\lambda/8$ was repeated by other groups [20,27]. New microspheres including Polystyrene [68] [28] and BaTiO$_3$ microspheres [29,30] [69] (immersed in liquid or solid encapsulated; mostly used particle now) were soon introduced and became widely used in the field,. Great efforts have since been devoted to the following areas: (1) scanning functionality development and complete super-resolution image construction. This is driven by the need of overcoming limitations of microsphere lenses whose imaging window is too small for many applications. (2) Mechanism exploration. Exact mechanisms have been debated in literature and considerable efforts were devoted to develop a complete theoretical model for the technique. (3) Development of next generation dielectric superlenses with improved resolution and imaging quality, new physics and fabrication method were explored. This has led to the development of all-dielectric metamaterial solid-immersion lens and nano hybrid lens. Because these lenses are metamaterial-based whose working mechanisms are completely different from previous counterparts, we would classify them as the 'second-generation particle-lens' and the previous microspheres the 'first-generation particle-lens'. (4) Easier access of superlens. Efforts also went to developing simpler version of superlens which layman or general public can access. Biological superlens based on naturally occurring spider silks was developed for this purpose. (5) New applications, including for example three-dimensional interferometry, endoscopy, and Lab-on-Chip (LoC) devices were developed.

In the following, fundamentals about optical super-resolution in microspheres and metamaterial particle-lenses will be firstly presented. The key progresses will then be reviewed, followed by the outlook of the field.

## 2 Super-resolution mechanisms: Photonic nanojet, Super-resonance and Evanescent jets

The super-resolution mechanisms for microsphere and metamaterial particle-lens are fundamentally different; they are thus presented separately below. Briefly speaking, for microspheres, the super-resolution is a combined result of several super-resolution effects, including photonic nanojet effect (weak super-resolution, non-resonant), super-resonance effect (strong super-resolution, resonant) and others including substrate effect and partial and inclined illumination effect, etc. For the second-generation metamaterial lens, strong super-resolution is a result of a new nanophotonic effect: nano-composite media made of high refractive index nanoparticles can effectively convert evanescent waves into propagating waves, and vice versa.

### 2.1 Microsphere and microcylinder

Optical resolution can be measured and calibrated by different means. In fluorescent imaging, 'point' source objects like fluorescent molecules are readily available. The system resolution can be precisely determined by measuring the point source function of the sources. In contrast, 'point' objects are rarely available in label-free imaging. To circumvent this issue, researchers often use finite size nanostructures and the resolution was measured based on observability of minimal discernible feature sizes which can be represented by the widths of the stripes, diameters of nanopores, edge-to-edge separations in dimers and clusters, etc. This has resulted in the resolution claims spanning the range from $\lambda/6$ to $\lambda/17$[31]. However, such claims were questioned and the resolution should be calibrated. A calibration procedure was suggested by Allen et. al. by convolution imaging object with a two-dimension PSF and compare with experimental results. The calibrated resolution for microsphere imaging is about $\sim\lambda/6$-$\lambda/8$ [32,19,27], instead of $\lambda/17$ as claimed by Lin [31]. It is important to note that a $\lambda/7$ resolution optical system can resolve $\lambda/25$ features, e.g. 15 nm gap in a bowtie sample as in Ref. [19]. In the following, we proceed to explain why a $\lambda/6$-$\lambda/8$ resolution is possible in microsphere-assisted imaging.

Despite of its simplicity, the underlying physical mechanism for microsphere imaging is quite complex. A complete theoretical model should consist of three elements: focusing of the incident light by microsphere, interaction of the incident light with the sample, and imaging of the scattered light in virtual mode. These processes are interlinked and coupled and complete model is still missing today. Most existing theoretical studies are based one or two processes above, leading to incomplete conclusions and debates [22,33,31,34,35,18,36,24,37,38,19,39-41]. At the beginning, Photon nanojet (PNJ) was considered as the main super-resolution mechanism [18,42]. However, it was soon realized that the resolution of PNJ (typically between $\lambda/2$ and $\lambda/3$ for n=1.5 microsphere) is insufficient to explain the experimentally observed resolution of $\sim\lambda/7$. Consequently, other mechanisms were discussed and investigated. Duan and co-workers reported that Whispery Gallery Mode (WGM) in microsphere can help boost imaging resolution up to $\sim\lambda/4$. Besides PNJ and WGM, a new physical effect, *super-resonance effect (SRE),* was recently discovered by Wang et.al. and Hoang et. al.

[43,44] independently .The SRE is caused by internal particle partial wave modes (not by scattering waves as in PNJ) and surface wave modes; it can generate super-resolution focusing with lateral resolution ~ λ/5 - λ/6. This new effect helps bridge the resolution gap between PNJ/WGM theories (~λ/4) and experiments (~λ/7). On the other hand, experimental resolution can be enhanced by substrate effect[45,33], partial and inclined illumination [27], microsphere partial immersion[46] and coherent illumination effect [19] as demonstrated by various authors. In the following, we briefly review the key properties of PNJ, SRE, broadband lighting source effect, substrate and partial and inclined illumination effect on super-resolution imaging. For more information on the basics of related theory (e.g. Mie theory) and PNJ (e.g. how PNJ evolves with respect to size and refractive index, medium effect, PNJ by other shaped particles, etc.), please refer to previous reviews written by us [36,33].

### 2.1.1) Photonic Nanojet

PNJ is a narrow, high-intensity electromagnetic beam that propagates into the background medium from the shadow-side surface of a plane-wave illuminated lossless dielectric microcylinder or microsphere of diameter greater than the illuminating wavelength, λ [19]. The PNJ is attained via light scattering by micro-sized particles cylinders, cubes, prisms and others, typical diameter between 1- 50 μm) [47,48] [49-57]. This is an important effect which is now widely used for subwavelength patterning, imaging, sensing, nanoparticle trapping and manipulation, etc. The earliest study relating to PNJ can be dated back to year 1999-2000 by one of the present author: Luk'yanchuk and co-worker first demonstrated that enhanced optical near-fields of a 500-nm silica sphere can be used as super-resolution lens (superlens) for subwavelength structuring of silicon surface[57]. Since then, there has been a strong continuing interest on the technique and many progresses have been made. Theoretical studies related to the topic also grew rapidly[58]. Without knowing the initial work by Luk'yanchuk et al., in 2004 Chen et al. coined the terminology PNJ -*'photonic nanojet'* [47], which is now widely adapted [59-70,47].

An example of PNJ focusing is shown in Fig. 2, where a 1μm-diameter Polystyne particle (n=1.6) in air is illuminated by a plane wave laser irradiation incident from the left (λ=248 nm), equivalent to size parameter q=2πa/λ=12.67, where a is radius of particle and λ the wavelength. The electric field is greatly enhanced in the near-field zone under the particle, but quickly decays along light propagation z-direction (almost exponentially as a character of near-field optics), from 59.6 at $z = a$ to 1.57 at $z = 2a$. Within the $z = a$ tangential plane, the distribution of laser intensity shows an elongated profile, whose long axis aligns with incident beam polarization while the super-resolution is observed in the cross direction. A resolution of 0.39 λ has been achieved, which surpasses the diffraction limit. The elliptical profile of near-field intensity has a physical root associated with the radial component of the electric field, $E_r$, which decays with $r$ as $E_r \propto 1/r^2$, in near-field zone. It quickly decays to zero in far field zone($r \geq \lambda$) [71]. In other words, scattering wave in the far field is transverse but contains both transverse and longitudinal components in near-fields.

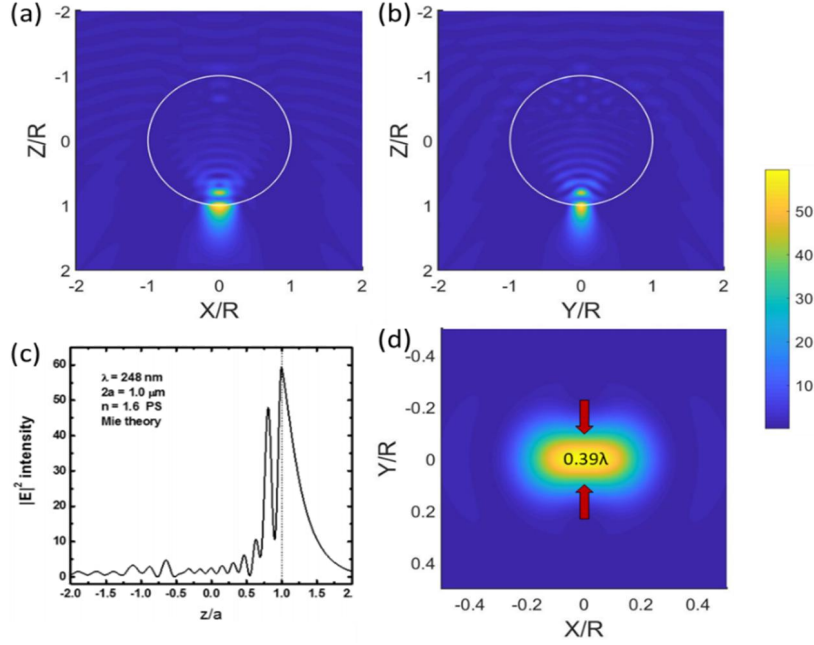

Fig. 2 Demonstration of near-field PNJ effect. Spatial electric field intensity distribution, $I = |E|^2$, inside and outside the 1.0 μm PS particle, illuminated by a laser pulse at $\lambda$ =248 nm (q=12.67), and (a) polarization parallel and (b) perpendicular to the image plane. The maximum intensity enhancement in calculations is about 60 for both regions. (c) Shows the intensity along $z$-axis. $z = 1.0$ is the position under the particle. (d) Super-resolution spot in at z=a, the tangent plane right under the particle.

The key properties of PNJ include [65,48]:

● The transverse beam width of the nanojet is usually just slightly better than 0.5λ, but in some parameter range it could reach limit close to $\lambda/2n$ [72], where n is the refractive index of particle. In the case of a glass particle with n=1.5, the resolution limit is about $\lambda/3 \approx 0.33\ \lambda$. PNJ focus is located outside of particle in the air zone which is accessible for nanoscale optical applications like nanopatterning and others.

● PNJ usually appears for diameter d of the dielectric microsphere or microcylinder from 2λ to more than 40λ if the refractive index contrast relative to the background medium is less than about 2:1.

● PNJ could maintain a subwavelength focusing jet along the path that can extend more than 2λ beyond the dielectric cylinder or sphere. This has led to a wide-spread but misleading claim that PNJ is a non-evanescent propagating beam within which evanescent wave doesn't contribute[73]. In fact, evanescent waves could play strong role in near-field PNJ as demonstrated in Fig. 2.

**2.1.2) Super-resonance effect (SRE)**

Besides PNJ, strong resonant modes can be excited in micro spherical cavity. For example, at n=1.5, q=26.94163, a super-resonance mode can be excited. We define 'super-resonance' as a giant field enhancement mode caused by microsphere's internal partial waves. In contrast, PNJ are induced by scattering waves outside the particle. Unlike WGM which can be excited in both microsphere and microcylinder, SRE only exist in microspheres. Fig. 3 demonstrates a typical SRE mode field distribution and its giant field enhancement amplitude.

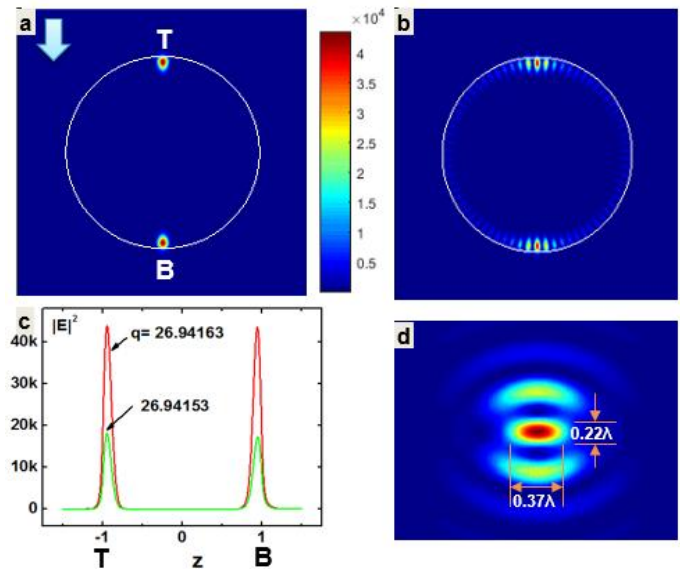

Fig. 3 Demostration of SRE. (a) Giant electric field intensity $|E|^2$ distribution in XZ plane and (e) YZ plane, for a linearly incident beam poiarized along x-direction. (c) Intensity across two hotspots (Points T, B), measured from top T to bottom B. (d) The FWHM spot size is 0.22λ (~λ/5, d) at position B.

In SRE mode, two near-symmetrical hotspots were generated near the top and bottom apex points (points T, B) of the particle (see Fig. 3a). The enhancement factor can be extremely large at these points, reaching the order of $10^4$ (peak value: 43,774), two orders higher than those in PNJ (typically on order of several tens or hundreds, Fig. 3c vs Fig. 1). Importantly, the SRE hotspot has a super resolution of 0.22λ which exceeds resolution limit of PNJ and WGM. The SRE is caused by the excitation of particular order of internal partial waves in Mie theory, e.g. L=43 for q=26.94163. It is very sensitive to the size parameter. For example, when q changes by 1e-4 from 26.94163 to 26.94153, the peak drops by more than 20,000 (see Fig. 3c). Details on SRE are outside the scope of this review and will be presented in a separate publication elsewhere. It is noted that Haong et al also reported super-resonance effect using same experimental parameters as in Fig.1 (a=2.37 um, n=1.46), they found that a ~0.24λ resolution can be achieved at wavelength 402.292 nm, in excess of solid immersion resolution limit of λ/ (2*1.46) =0.34λ)[43]. We believe SRE is vital to microsphere super-resolution imaging.

### 2.1.3) Broadband illumination effect

In microsphere imaging, theoretical analysis of white light interaction with microsphere is often simplified by using its peak wavelength (e.g. 550 nm or 600nm). This causes the broadband nature of illumination source being neglected. In fact, the application of white light will lead to the simultaneous excitation of various intensity field patterns across a size parameter range. For example, for a 3-μm-diameter particle under white lights, its size parameter q will range from 13.46 (λ = 700 nm) to 23.56 (λ = 400 nm). Using Mie theory, we calculated the XZ-plane $|E|^2$ field distribution for varying q with step size accuracy $\Delta q = 0.1$ for both spheres and cylinder, as shown in Fig.4. For spheres, it shows the co-existence of at least three different field modes or patterns, i.e. SRM at q = 18.4 (Fig. 4a), a typical PNJ at q=20.9 (Fig. 4b) and WGM at q = 21 (Fig. 4c)when excited by white light. For cylinder, the results are slightly different, we didn't observe the super-resonance modes in cylinders but see evidence of more pronounced WGM mode q = 19.2 (Fig. 4e). See online movies [74,75] for details on light focusing by an n=1.5 particle and cylinder when size parameter increases varies from 0 to 600. The flicking phenomena in the movie indicate the modes swapping between PNJ, WGM and SRM. In a complete theoretical model, one should consider spectrum profile of lighting source and integrate all modes contribution.

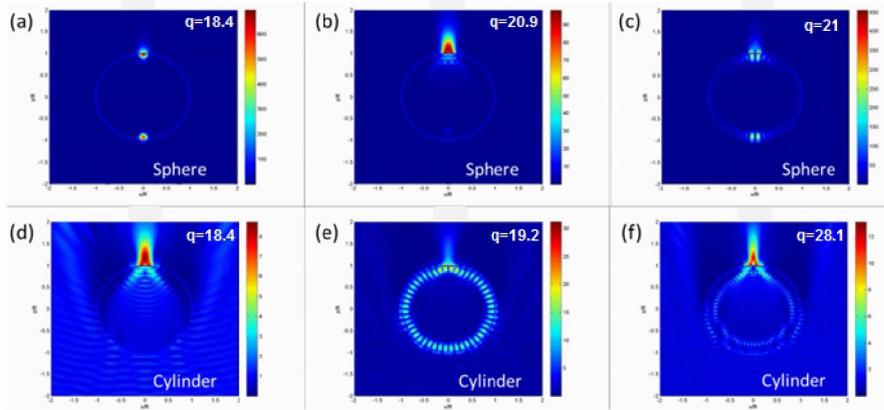

Fig. 4 Typical $|\mathbf{E}|^2$ field distributions of (a-c) spheres and cylinders (d-f) with refractive index $n = 1.5$ at varying size parameter q with calculation step size $q = 0.1$. (a) super-resonance mode of sphere, (b) usual jet mode of sphere, (c) whispery gallery mode of sphere (d) usual jet mode of cylinder, (e) strong whispery gallery mode of cylinder (f) weakly excited whispery gallery mode of cylinder. See supplementary video for details.

## 2.1.4) Substrate effect

Substrate effect is clearly observed in experiments, i.e. imaging magnification and resolution vary for different samples. This is, however, less studied and understood. Sundaram and co-workers performed imaging analysis using full-wave FEM simulation of the light propagation from the target through a microsphere, an objective lens and then to the imaging plane, as shown in Fig. 5 [45]. The results show that longitudinally polarized dipole can achieve better resolution than transverse dipole (Fig. 5b and c). More importantly, it was shown that substrate plays an important role in imaging magnification and resolution. For example, as shown in Fig. 5d, imaging resolution for air (without substrate), aluminium oxide ($Al_2O_3$) and fused silica ($SiO_2$) are $0.28\lambda$, $0.24\lambda$ and $0.26\lambda$, and magnification 6, 6.6 and 6.4, respectively. Another numerical simulation performed by us, as shown in Fig. 5e, reveals complex nature of high spatial frequency evanescent wave decoupling process from sample surface[33]. A parallel-to-surface wave vector, marked by the red solid line, was scattered by the particle, converting into propagating beam transferring to the far-field (red dashed line).

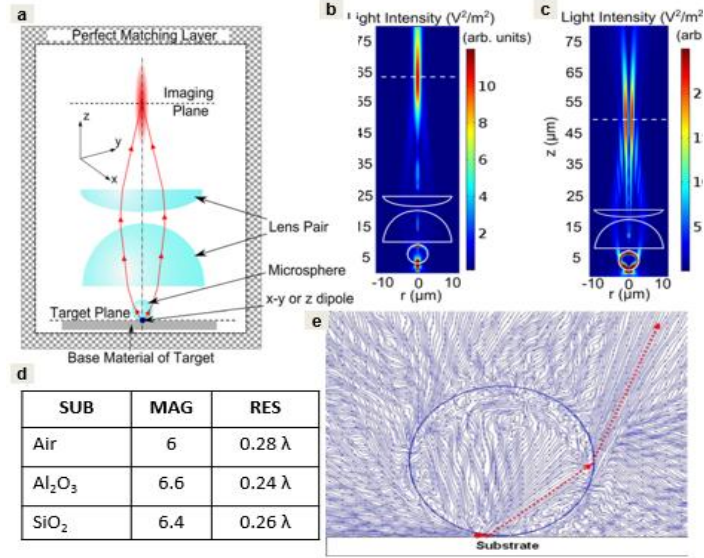

Fig. 5 (a) Schematic of the simulation setup consisting of the substrate target, microsphere, and lens pair. (b) Light intensity ($|E|^2$) distribution induced by a transverse dipole (x-y dipole), and (c) a longitudinal dipole (z-dipole), in air with a microsphere lens of diameter 6 $\mu$m and refractive index n=1.4. The dashed lines indicate the imaging plane. (d) Table of magnification and resolution for different substrates. SUB: Substrate, MAG: magnification, RES: resolution. (e) Numerically simulated point dipole parallel-to-surface evanescent wavevector decoupling process, from near-field to far-field.

## 2.2 Dielectric nanoparticle-based metamaterial superlens

As many studies reveal that most metal-based metamaterials suffer from intrinsic loss at high optical frequencies, which cause them to be unfavorable in near-infrared and visible region [76-79]. Moreover, metal-based plasmonic components have low transmission efficiency at optical frequencies, thus making them less useful for optical waveguiding over long distances or through bulk three-dimensional (3D) structures[80] [81]. Unlike the metal-based metamaterials, dielectric metamaterials use the near-field coupling between transparent (low absorption), high-refractive index dielectric building blocks, which perform similar optical phenomena to metallic nano-resonators, but with much lower energy dissipation[82-85]. Meanwhile, high transmission and diffraction efficiencies of dielectric components make it possible to move optical metamaterials from current 2D metasurfaces or layered metamaterials to truly 3D metamaterials[80]. New types of particle-lens - metamaterial solid immersion lens (mSIL)- has recently been proposed and developed by us. This is based on metamaterial concept and a new super-resolution mechanism. The mSIL is a 3D metamaterial formed by densely stacking of high-index TiO$_2$ nanoparticles (Fig. 6a). Here, the reso-lution is determined by the particle size instead of wavelength. This new nanophotonic effect is shown in Fig. 6 by a full-wave numerical simulation.

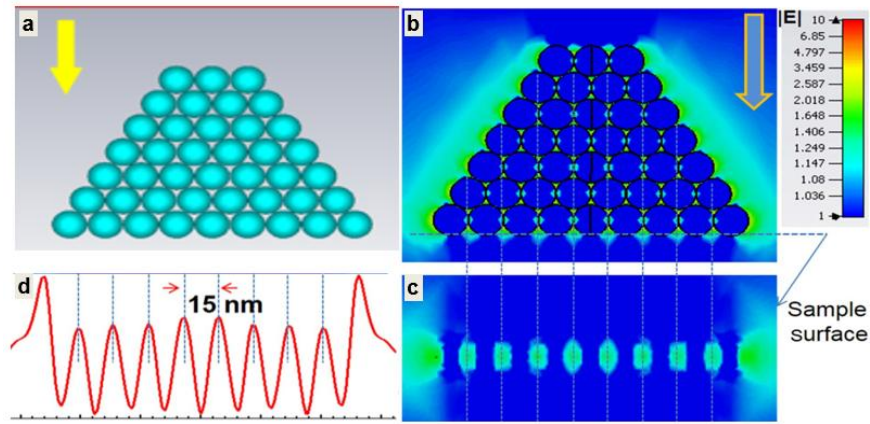

Fig. 6 Propagating wave scattering by densely packed all-dielectric nanoparticles medium. (a) schematic drawing of a nanoparticle stacked media. (b) Plane wave ($\lambda$= 550 nm) passing through the stacked TiO$_2$ nanoparticles. Electric field hotspots are generated in the gaps between contacting particles, which guides light to the underlying sample. (c) Large-area nanoscale evanescent wave illumination can be focused onto the sample surface because of the excitation of nanogap mode. (d) The size of the illumination spots is equal to the particle size of 15 nm.

The basic simulation structure of the artificial media is a closely stacked 15-nm anatase TiO$_2$ nanoparticle composite, in which tiny air gaps between the particles exist, resulting in a dense scattering media. Fig. 6 shows the simulation results of electric field distribution in the media when applying a plane wave illumination at a wavelength of 550 nm from the far field. Electric field confinements are observed in the gaps between nanoparticles, indicating the ability of the composited media to modulate and confine visible light at the nanoscale, as shown in Fig. 6 (b). Since TiO$_2$ is nearly free of energy dissipation at visible wavelengths, this near-field coupling effect among neighboring nanoparticles can effectively propagate through the media over long distances, forming an arrayed "patterned illumination" landscape on the surface of an underlying substrate, as shown in Fig. 6 (c). These illumination spots are evanescent in nature, containing high–spatial frequency components. Their sizes are mainly determined by the size of TiO$_2$ nanoparticles, having a full width at half maximum (FWHM) resolution of ~8 nm, as shown in Fig. 6 (d). Therefore, it is expected that nanoparticle composited media will have the unusual ability to transform the far-field illumination into large-area nanoscale evanescent wave illumination focused on the object surface within the nearfield region. This novel nanophotonic effect is somewhat similar to that of aperture near-field scanning optical microscopy (NSOM), in which evanescent wave illumination is transmitted from the subwavelength aperture at the tip of a metal-coated optical fiber, and the size of the illumination spot is not limited by the incident wavelength but by the aperture size[86,87]. However, the single nanoaperture design in NSOM suffers from several limitations, such as low optical throughput, long scanning time, and insufficient contact between aperture probe and object surface. In our design, the array of TiO$_2$ nanoparticles can act as thousands of near-field probes to simultaneously illuminate the sample surface, and the strength of the focused evanescent wave illumination can be maximized owing to the near perfect solid immersion of imaging object by TiO$_2$ nanoparticles. Moreover, theoretically, the size of the evanescent wave illumination spots can be further reduced by using smaller anatase TiO$_2$ nanoparticles or higher refractive index rutile (n = 2.70) TiO$_2$ nanoparticles, and this near-field illumination may also be useful for other applications such as nanoscale light harvesting and sensing.

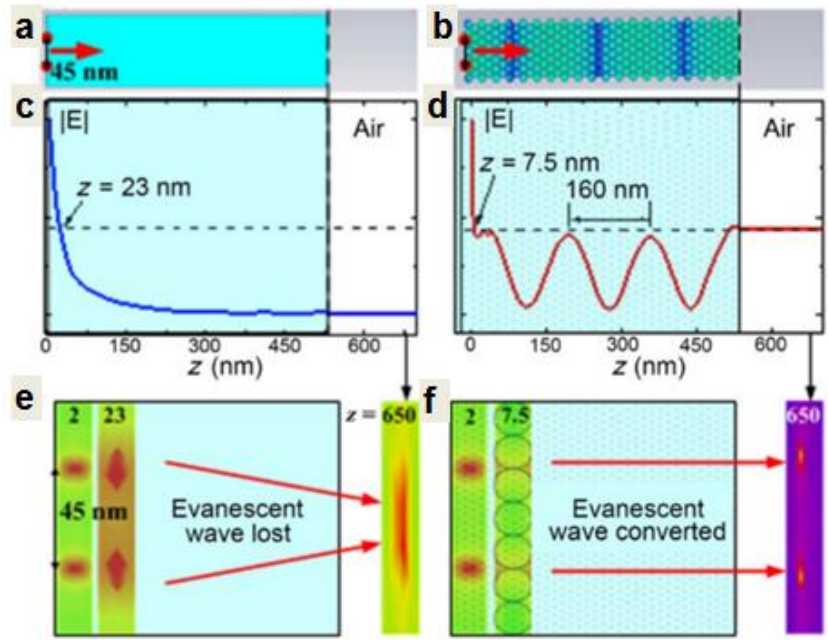

Fig. 7 Comparisons between homogeneous and nanoparticle composited media. (a) Mean electric field amplitude as a function of distance from point sources (y-polarised, incoherent). The amplitude decays exponentially. Most evanescent wave energy was lost within 50 nm distance. (b) In nanoparticle composited material, evanescent waves interact with TiO$_2$ nanoparticles and turn into propagation waves which travel outward to far-field. A periodicity of 160 nm was observed. (c-f) Two-point sources (45 nm separation) imaged with (c) homogeneous and (f) composited material, at positions z=2 nm (near source), Z=23 nm (near-field, inside slab) and Z=650 nm (far-field, outside slab). In far-field, (e) the homogenous media fails to resolve the two points while (f) the composited media can successfully resolve them.

According to the reciprocal principle [88], the conversion process in nanoparticle composited media (Fig. 7) from propagating waves to evanescent waves can be optically reversed. In other words, an array of evanescent wave source located on the bottom surface of the nanoparticles composited media will be converted back into propagating waves by the media. This is confirmed by two–point source calculation. For comparison purposes, both conventional media (homogeneous anatase TiO$_2$ material) and our metamaterial media (stacked 15 nm anatase $TiO_2$ nanoparticle) were simulated. Using slab geometry, we demonstrate in Fig. 7 that evanescent waves behave differently when interacting with conventional media, Fig. 7 (a), and composited metamaterial media, Fig. 7 (b and d). As shown in Fig. 7 (c), in homogeneous media the evanescent waves decay exponentially as expected when the distance to sources increases. The loss of evanescent waves causes reduced resolution. Fig. 7 (e) demonstrates the inability of conventional media to resolve two-point sources (separated by 45 nm) in far-field zone (*e.g.* $z$=650 nm > $\lambda$=550nm). The two points are only resolvable in near-fields when distance to source is extremely small, typically smaller than 50 nm in present case. In nanoparticle composited media, however, the evanescent waves experience strong interaction with $TiO_2$ nanoparticles, which causes effective conversion of evanescent waves into propagation waves. The converted waves are mainly guided through the gaps between particles. As shown in Fig. 7 (d), electric field received at far-field region ($z$>$\lambda$) is about |E|~0.45 (since source amplitude |E|=1, this corresponds to $|E|^2$= 20% of total evanescent energy), which is comparable to the field strength at $z$=7.5nm (near-field). This means near-field energies are indeed converted and transported to the far-field. It is also interesting to notice the periodic modulation effects of E-field inside the composited media, which can be seen as a signature of this design. Since there is no material loss in composited media, the periodic propagation experiences an un-damped modulation, showing an effective period of 160 nm. These waves propagate outwards effectively from near-fields into far-fields and contribute to super-resolution. In Fig. 7 (f), it is demonstrated that at far-field (*e.g.* $z$=650nm), the two-point sources are reconstructed and clearly discernible. Comparing to metal-based superlens and hyperlenses whose resolution is limited by material losses in metal[89,9] [90,91], the proposed all-dielectric nanoparticles media is free from loss problems; its resolution is mainly affected by the excitation of evanescent waves and the conversion efficiency of evanescent waves into propagating waves, as well as the effective capture of the subwavelength information in the far field. This makes it possible to design a perfect imaging device using dielectric nanoparticles as building block. In section below, we will show our experimental demonstration of a fabricated $TiO_2$ nanoparticle-based metamaterial solid immersion lens (mSIL) and its super-resolution imaging performance (45 nm).

# 3 Microsphere nanoscopy: key progresses

In this section, we review key progress in the field, including microsphere-assisted confocal microscopy, integrated superlensing microscope objective, scanning imaging, holography and endoscopy

## 3.1 Confocal imaging and resolution quantification

In widefield, the imaging contrasts are often low and unsatisfactory due to the presence of out-of-focus light in the final image. To enhance the contrast, efforts were required to optimize the microscope lighting and image capturing settings. In contrast, confocal microscopy generally has much better optical contrast and improved resolution; this is achieved by placing a tiny pinhole before the detector to eliminate the out-of-focus light in the final image. Several groups evaluated the imaging performance of microsphere-assisted confocal imaging by integrating microsphere superlens with a 405 nm laser scanning confocal microscope (e.g. Olympus OLS4000). Wang et al. imaged 140 nm nanolines separated 40 nm away using PS particle in air, and emphasised the importance of using single isolated particle in confocal imaging where artefacts could arise when two neighbouring microspheres interferences with each other [33]. Yan et al. imaged 136 nm gold nanodots separated 25 nm away, using PS and fused silica particle in air as well [31]. Using BTG particle, Allen and co-workers imaged 120 nm gold nanocylinders separated 60 nm away (Fig. 8a) and a 185 nm nanobowtie separated 15 nm away (Fig. 8e) [19]. They also argued that the resolution based on observation of minimal discernible features can lead to misleading results if, for example, interpreting the 15 nm gap in nanobowtie (Fig. 8e) would imply the resolution over λ/27. They suggested an approach to compute the resolution based on the analogy with the classical theory where the image, $I(x,y)$, is considered as a convolution of a point spread function (PSF) and the object's intensity distribution function, $O(u,v)$. This can be expressed in the standard integral form[19]:

$$I(x,y) = \int_{-\infty}^{\infty} \int_{-\infty}^{\infty} O(u,v) \, PSF\left(\frac{x}{M} - u, \frac{y}{M} - v\right) du \, dv \quad , \qquad (1)$$

in which the integration is performed in the object plane where the coordinates $(x_o, y_o)$ are linearly related to the image plane via the magnification M as: $(x_o, y_o) = (x_i/M, y_i/M)$. A Gaussian function for $PSF(x_o, y_o)$ with the full width at half maximum (FWHM) being a fitting parameter. Based on the Houston criterion, fitted values of FWHM in the object plane were considered as resolution of the system. Based on this, the image reconstruction with the λ/5.5 PSF resolution allows obtaining a high-quality fit to the experimental results, as illustrated in Figs. 8(g) and 8(h). This means that the resolution for 15 nm gap sample in Fig. 8e is λ/5.5, instead of λ/27. Similarly, it was shown the 60 nm gap sample in Fig. 8a is λ/7 (Fig 8c and d).

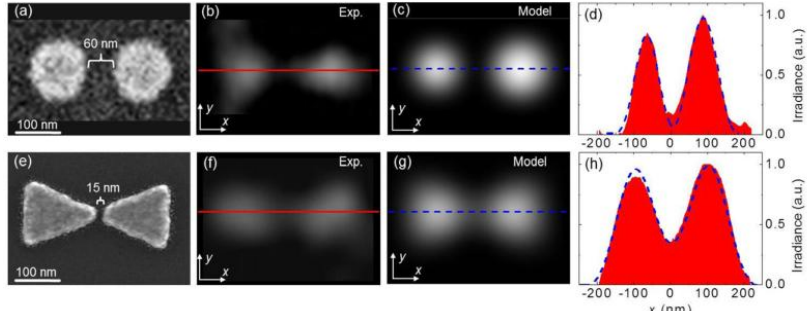

Fig. 8 SEM image of the Au dimer and bowtie, respectively. The dimer is formed by 120 nm nanocylinders with 60 nm separation. (b,f) confocal Images of dimer and bowtie in (a,e) obtained through the 5 μm and 53 μm BTG microsphere, respectively. (c,g) Calculated images by convolution method at PSF resolution of λ/7 and λ/5.5, respectively. (d,h) Comparison of the measured (red background) and calculated (dashed blue curves) irradiance profles through the cross-section of the Au dimers (x-axis).

### 3.2 Scanning microsphere super-resolution imaging

Microsphere has a viewing window around a quartz size of its diameter. Considering particles with diameter 1-100 um, the viewing window is less than 25 um which is too small for many practical applications. Controlling the position of the microspheres is required for generating a complete image of sample. Scanning functionalities are, hence, highly sought by the researchers working the field. There are several proposals existing in the literature, including for example the microsphere-embedded coverslip, superlensing microscope objective, swimming lens and AFM-style scanning microsphere-lens.

### 3.2.1 Microsphere-embedded coverslip

One of the earliest but important developments is the use of liquid-immersed BTG microspheres for super-resolution imaging (compatible with bio-samples). This configuration is still widely used today, with resolution between λ/4 -λ/7 depending on the particle size, as demonstrated experimentally [69] [29]. The liquid-immersion design was soon extended to solid-immersion design in a coverslip form [92,19,93], as shown in Fig. 9(a). The fabrication steps are illustrated in Fig. 9(f): first, BTG spheres were spread on a flat surface. Second, a liquid PDMS (or other resins) was casted and cured, producing a solid thin film coverslip whose thickness can be varied from hundred micrometres to millimetres. Finally, the coverslip film was separated and used for imaging experiments. The coverslip can be attached to a moving stage (Fig. 9a) for performing super-resolution imaging at specific site. The motion of the coverslip can be provided by a metallic probe inserted in PDMS and connected to a micromanipulator (Fig. 9b). Fig. 9 (c-e) shows an example of translation of coverslip assisted by IPA lubricant for imaging. The imaging quality and stability are both improved by using coverslip design.

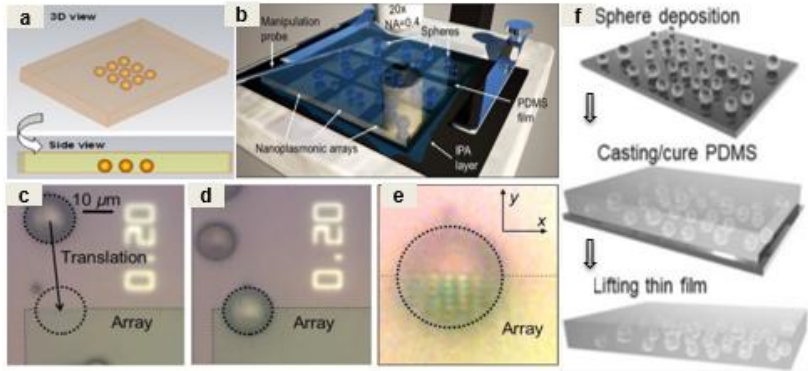

Fig. 9 (a) Microsphere-embedded coverslip design (b) setup and translation of the coverslip lubricated with IPA. (c) the embedded BTG sphere is ~40 μm away from the edge of an Au nanoplasmonic array. (d) The same sphere is at the border of array. (e) By changing the depth of focus, the dimers are seen near the array's edge. (f) coverslip fabrication steps, sphere deposition on a flat surface, PDMS casting and curing, lifting thin film to obtain the coverslip.

### 3.2.2 Superlensing microscope objective

The coverslip lens discussed above can be manually manipulated in a way like classical coverslip, offering the freedom to position particle-lens at desired location. Scanning of microsphere superlens, however, requires synchronisation with microscope objective for full super-resolution image construction. We proposed an improved design which solves the synchronisation problem of coverslip superlens and objective lens. The idea is to use a custom-made lens adaptor to integrate these two lenses to form a superlensing microscope objective lens.

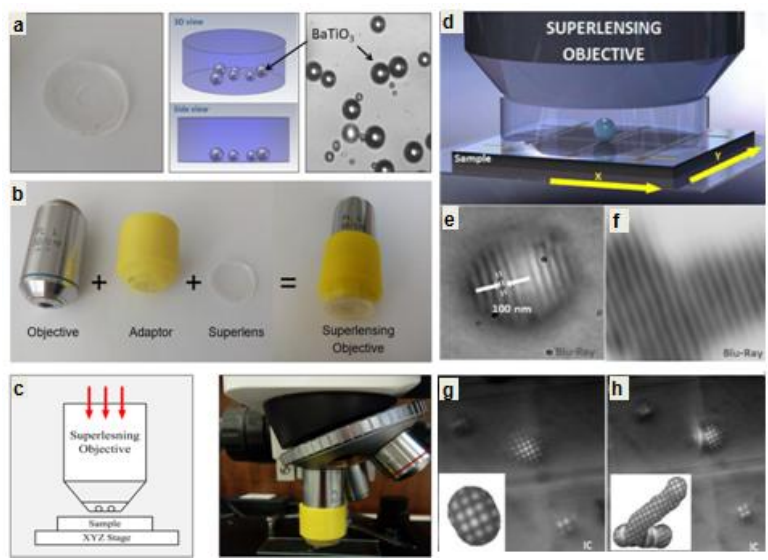

Fig. 10 Superlensing objective lens. (a) The BTG superlens was fabricated by encapsulating a monolayer of BTG microsphere inside a PDMS material. (b) the super objective was made by integrating a conventional microscope objective lens (e.g. 50x, NA: 0.70, or 100x, NA:0.95) with a BTG microsphere superlens using a 3D printed adaptor (c) Experimental configuration for super-resolution imaging using developed objective which was fitted onto a standard white light optical microscope. (d) Scanning illustration. (e) Single particle imaging. Of Blu-ray (f) Scanning generated full Blu-ray image. (g) Nanochip imaging (f) scanning nanochip imaging.

The key concept and design of superlensing microscope objective is illustrated in Fig. 10(b) and (c). A conventional microscope objective (OB) lens with magnification factor between 40x and 100x, NA between 0.7 and 0.95 was selected. A lens adaptor was designed in CAD software (e.g. Solidworks) and then printed with a 3D plastic printer. The adaptor has a tube size fit to the objective lens tube, with reasonable frication allowing up-down adjustment. A coverslip superlens (Fig. 10a) was bounded to the bottom end of the adaptor using high-adhesive glue. This results in an integrated objective lens consisting of conventional OB and a Coverslip Microsphere Superlens (CMS). The imaging resolution will be determined by the coverslip superlens while the conventional objective lens provides necessary condition for illumination. The obtained superlensing lens can be easily fitted to any existing conventional microscopes; The scanning was performed using a high-resolution nano-stage. In experiments, the superlensing objective lens was kept static and the underlying nano-stage moves and scans the samples across the objective lens. The imaging process was video recorded using a high-resolution camera and images were constructed from the video. Fig.11 (e, f) and (g,h) demonstrates the scanning super-resolution imaging by the integrated objective lens, which makes large-area super-resolution imaging possible.

### 3.2.3 Swimming lens

Li et al. designed a "swimming lens" technique in which the microsphere lens was propelled and scanned across the sample surface by chemical reaction force in a liquid[74]. This approach enables large-area, parallel and non-destructive scanning with sub-diffraction resolution. Fig. 11 illustrates the schematic diagram and scanning imaging process.

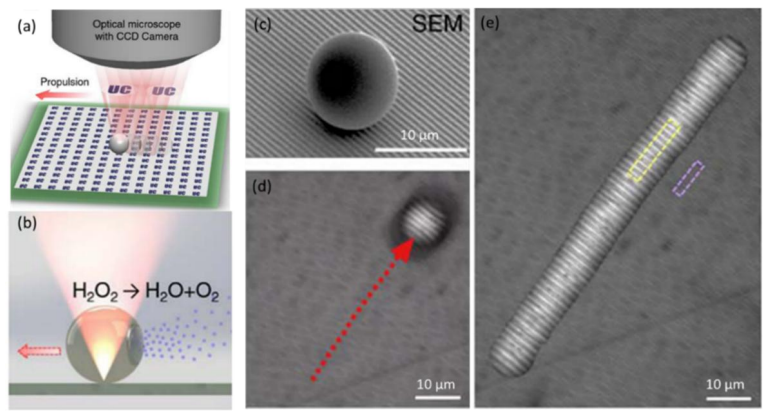

Fig. 11 Swimming lens design. (a) Schematic illustration. (b) Schematic illustration of the chemically powered propulsion and light illumination through the microsphere. (c) SEM of 10 μm PS microsphere on a 320 nm grating structure. (d) Microsphere imaging and arrow shows the scanning motion. (e) Reconstructed image by stitching from individual video frame[74].

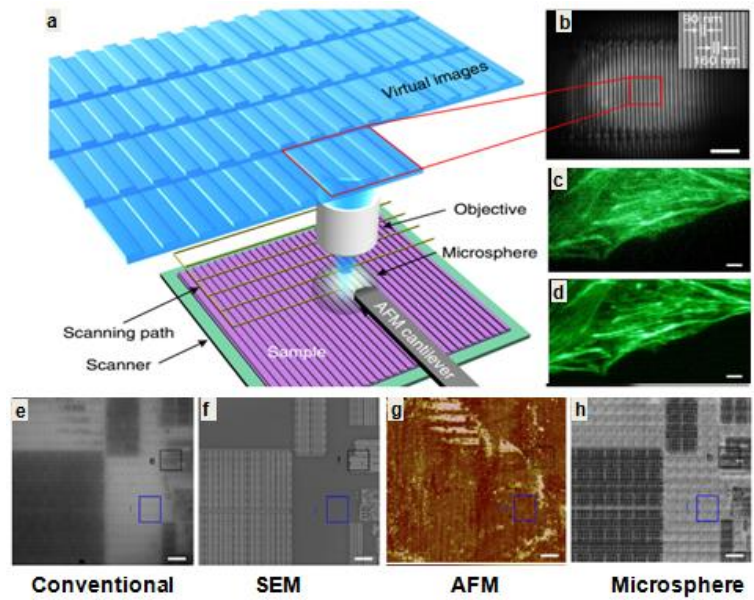

Fig. 12 AFM-style scanning microsphere imaging system. (a) Schematic of the construction of a microsphere-based Microsphere-based scanning superlens microscopy (SSUM) that integrates a microsphere superlens into an AFM scanning system by attaching the microsphere to an AFM cantilever. The objective picks up the virtual images containing sub-diffraction-limited object information and simultaneously focuses and collects the laser beam used in the cantilever deflection detection system. (b) An original virtual image observed using the microsphere superlens. The inset shows an SEM image. (c, d) Backside and frontside images, respectively, of the AFM cantilever with an attached microsphere superlens. Scale bars, 2 μm (b); 50 μm (c, d)[75].

### 3.2.4 AFM-style scanning microsphere lens

Most recently, Wang et al. introduced a non-invasive, environmentally compatible and high-throughput imaging technique called scanning superlens microscopy (SSUM), which applying the AFM principle by attaching microsphere onto AFM tip for scanning imaging (Fig. 12)[75]. This system has high precision in maintaining distance between microsphere and the objects. It has capabilities of operating in contact scanning mode and constant-height scanning mode, therefore variety of samples such as stiff sample and sensitive specimens can be efficiently imaged. Besides, Krivitsky et al. attached a fine glass micropipette to the microsphere lens to scan the particle[73]. The method allows precisely controlling the microsphere position in three dimensions as well.

### 3.3 Endoscopy application:

In vitro nano-scale imaging by super-resolution fluorescence microscopy techniques, such as STED, STORM and PALM has achieved significant development in recent years. However, in-vivo endoscopy observation of deep and dense tissues inside body has limited by its optical resolution at ~1 µm. Wang and co-workers developed a new endoscopy method by functionalizing graded-index (GRIN) lens with microspheres (30-100µm BTG particle) for real-time white-light or fluorescent super-resolution imaging[37]. The capability of resolving objects with feature size of ~λ/5, which breaks the diffraction barrier of traditional GRIN lens-based endoscopes by a factor of two, has been demonstrated by using this superresolution endoscopy method. Fig. 13 shows the experimental step and imaging test results. The convectional endoscope cannot resolve 100-nm sized fluorescent nanoparticles while new endoscope containing microspheres can clearly resolve them in super-resolution. The super-resolution strength could be further improved by new metamaterial particle-lenses with better resolution. Further development of such a superresolution endoscopy technique may provide unprecedent new opportunities for in vivo diagnostics and therapy as well as other life sciences studies.

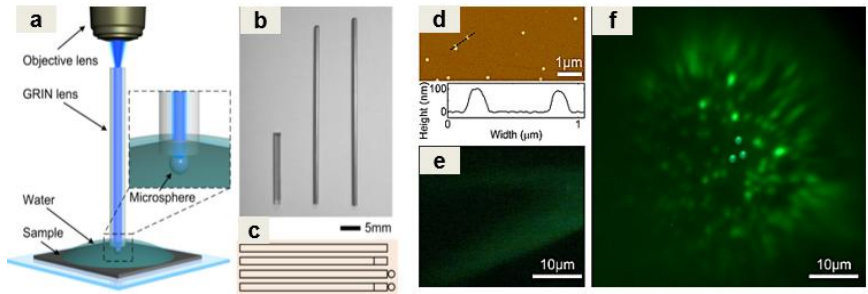

Fig. 13 Microsphere Super-resolution endoscopy. (a) Schematic of microsphere super-resolution endoscope. (b) Three GRIN lens based endoscope probes, a 1.8 mm diameter singlet (NA = 0.46), a 1 mm diameter singlet (NA= 0.11) and a 1 mm diameter compound doublet (objective lens NA = 0.42, relay lens NA =0.11). The second and third probes are covered by protective metal sleeves. (c) Schematic of singlet, doublet and microsphere-functionalized singlet or doublet super-resolution endoscopes. (d) AFM image of 100 nm fluorescent nanoparticle. (e ) Fluorescent image obtained by using a conventional doublet endoscopy probe. (f) super-resolution image obtained using microsphere (80µm BTG particle) super endoscope.

### 3.4 3D interferometry and optical profiling:

3D super-resolution profiling and imaging was recently demonstrated by integrating a microsphere with a white-light Interferometry system (Linnik [38] or Mirau[94]).

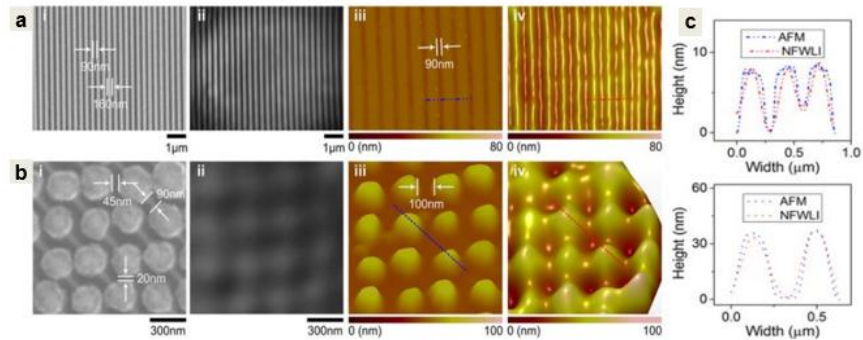

Fig. 14 Near-field assisted white-light interferometry (NFWLI). (a) 90 nm features (b) 20-90 nm features in a CPU chip. (i) SEM images. (ii) virtual images generated by 69 μm and 19 μm BTG microspheres in (a) and (b), respectively. (iii) AFM images. (iv) NFWLI images. ( c) comparisons of the cross-sections marked by lines in (a(iii,iv)–b(iii,iv)), respectively.

This near-field assisted white-light interferometry (NFWLI) method takes advantage of topography acquisition using white-light interferometry and lateral super-resolution enhancement by microsphere. The ability to discern structures in central processing units (CPUs) with minimum feature sizes of approximately 50 nm in the lateral dimensions and approximately 10 nm in the axial dimension within 25 s (40 times faster than atomic force microscopes) was demonstrated by Wang and co-workers[38], as shown in Fig. 14.

The ability of discerning structures in super-resolution along vertical direction could be explained by the conversion of evanescent waves into propagating waves mediated by the microsphere; the converted beam reaching the far-field interferences with reference beam and produce super-resolution vertical profiling. The technique has great potential for applications in fields where fluorescence technology cannot be utilized and where detection of nanoscale structures with a large aspect ratio is needed, i.e., in cases where the tip of a scanning probe microscope cannot reach.

## 4 Metamaterial dielectric superlens

Based on metamaterial concept and new super-resolution mechanisms, two types of metamaterial dielectric superlenses were developed, i.e. metamaterial solid immersion lens (mSIL) and nanohybrid lens (nHL).

### 4.1 mSIL

A new "nano–solid-fluid assembly (NSF)" method was developed by using 15-nm $TiO_2$ nanoparticles as building blocks to fabricate the three-dimensional (3D) all-dielectric metamaterial at visible frequencies (see Fig. 15). As for its optical transparency, high refractive index, and deep-subwavelength structures, this 3D all-dielectric metamaterial-based solid immersion lens (mSIL) can produce a sharp image with a super-resolution of at least 45 nm under a white-light optical microscope, significantly exceeding the classical diffraction limit and previous near-field imaging techniques.

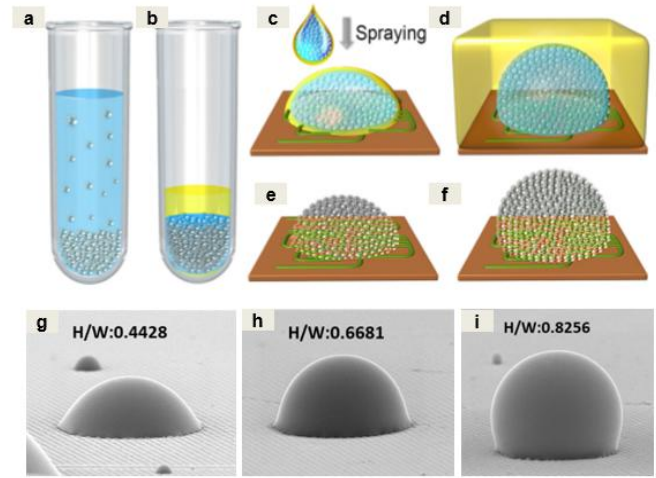

Fig. 15 Nano-solid-fluid (NSF) assembly method for fabrication of mSIL. (a) Anatase TiO2 nanoparticles were centrifuged into a tightly packed precipitate. (b) The supernatant was replaced by an organic solvent mixture consisting of hexane and tetrachloroethylene to form a TiO$_2$ NSF. (c) To prepare a hemispherical mSIL, the NSF was directly sprayed onto the sample surface. (d) To prepare a superhemispherical mSIL, the NSF was sprayed onto the sample surface covered by a thin layer of organic solvent mixture. (e, f) After evaporation of the solvents, the nanoparticles underwent a phase transition to form a more densely packed structure of mSIL. (g,h,i) different height to width (H/W) ratio mSIL.

The super-resolution imaging performance of mSIL was illustrated in Fig. 16. The magnification factor sharply increases with the height-to-width ratio of the mSIL approaching unity (a spherical shape) and reaches 5.3 at a height to-width ratio of 0.82. With the further increase of this ratio, the contrast of the virtual image gradually disappears. Using geometrical ray tracing, we theoretically fit the experimental magnification curve and inversely derived that the mSIL media has an effective index of 1.95 and a high particle volume fraction of 61.3%. This packing fraction is close to the random close packing limit (~64%) of monodisperse hard spheres (46), indicating an intimate contact between TiO2 nanoparticles throughout the media. Moreover, the mSIL presented here exhibits a wide field of view, which is approximately linearly proportional to the width of the mSIL. In (c, d) and (e, f), super-resolution imaging was demonstrated; both 60 nm and 45 nm features can be clearly resolved with mSIL. The imaging quality, compared to 1<sup>st</sup> generation microsphere lens, has been greatly improved and being consistent when imaging different types of samples, no matter it is metal, semiconductor or dielectric samples. This indicates the new physical mechanism is robust and dominant in mSIL imaging. Fig. 17 shows another successful example of mSIL unpublished previously, where 60 nm diameter Polystyrene nanoparticles were clearly imaged by the mSIL, directly under white light. In contrast, microsphere lenses fail to resolve these particles.

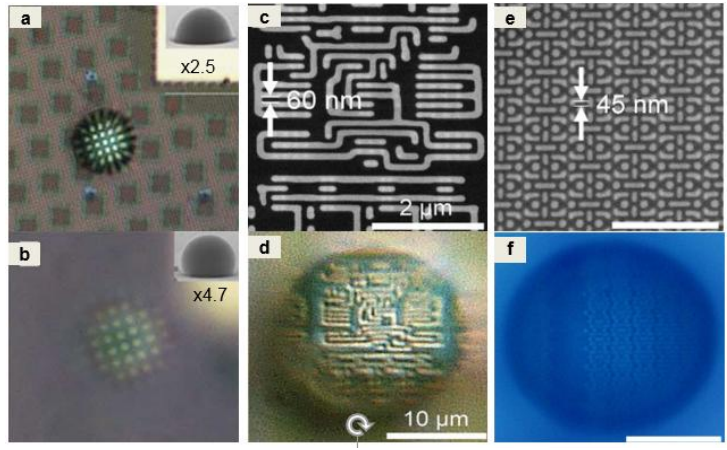

Fig. 16 Super-resolution imaging performance of mSIL. (a) Magnified image (x2.5 times) by a mSIL shown in inset, (b) Magnfication factor of x4.7 for a larger heigh/width ratio mSIL. (c) SEM (d) mSIL images of 60 nm features. (e) SEM and (f) mSIL images of 45 nm features.

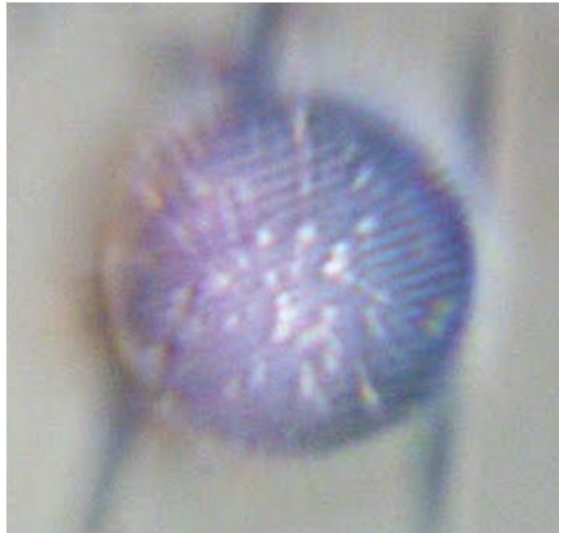

Fig. 17 Direct super-resolution of 60-nm-diameter PS nanosphere deposited on a grating sample surface.

The NSFA method we have described is simple and versatile and can be readily extended to assemble TiO2 nanoparticles or even other dielectric nanoparticles into arbitrarily shaped metamaterial-based photonic devices (for example, a TiO2 wire used as an optical fiber; see fig. s9 in [95]). Further combining techniques such as nano-imprinting and nanofluidics may lead to compact and inexpensive nanophotonic devices for cloaking, optical interconnect networks, near-field sensing, solar energy utilization, etc.

### 4.2 Nanohybrid lens:

Nanohybrid lens is a new class of metamaterial lens with controllable refractive index. This is achieved by uniformly distributing high refractive-index (RI) $ZrO_2$ material into polystyrene matrix. A nanoparticle-hybrid suspension polymerisation approach was developed for such purpose; high-quality microspheres ZrO2/PS hybrid microspheres have been synthesized with highly controllability in shape and refractive index (np = 1.590–1.685) [96]. Fig. 18 ZrO2/PS nanohybrid particle-lens. (a) SEM image and SEM mapping photographs of (b) Zr, (c )C and (d) O in the nanohybrid microspheres. (e) Schematic of the dimethyl-silicone semi-immersed microsphere for super-resolution imaging. (f) Nanochip sample with 75 nm and 60 nm features. (g) imaging by n=1.590 particle. (h) imaging by n=1.685 particle. shows nanohybrid particle-lens (a-d) and its super-resolution imaging performance configured in a semi-immersed liquid environment (e). Comparing (g) and (h), it is obvious that increasing the refractive index of microspheres from 1.590 to 1.685 improve both the imaging resolution and quality. A 60 nm resolution has been obtained in the widefield imaging mode and a 50 nm resolution in the confocal mode. The synthesis of hybrid microspheres is feasible, easily repeatable and designable in shape, size and refractive index. Accordingly, the approach is quite general, and can be readily extended to prepare a series of hybrid microspheres with various refractive index and optical transparency, by changing the types of polymers and nanoparticles. The as-synthesized nanohybrid colloidal microspheres can be used not only in optical nanoscope for super-resolution imaging with visible frequency, but also in some potential fields of nanolithography, optical memory storage, and optical nano-sensing.

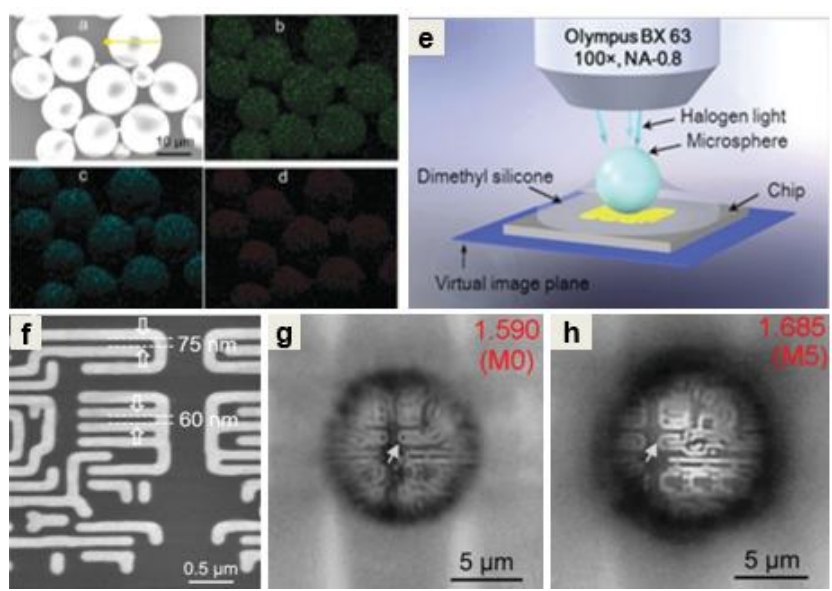

Fig. 18  ZrO$_2$/PS nanohybrid particle-lens. (a) SEM image and SEM mapping photographs of (b) Zr,  (c )C  and (d) O in the nanohybrid microspheres. (e) Schematic of the dimethyl-silicone semi-immersed microsphere for super-resolution imaging. (f) Nanochip sample with 75 nm and 60 nm features. (g) imaging by n=1.590 particle. (h) imaging by n=1.685 particle.

## 5. Biological Superlens

Fabrication of superlenses  is often complex and requires sophisticated engineering processes. Clearly an easier model candidate, such as a  naturally occurring superlens, is highly desirable. Recently, Monks et al. reported a biological superlens provided by nature: the minor ampullate spider silk spun from the Nephila spider[97]. This natural biosuperlens can distinctly resolve 100 nm features under a conventional white-light microscope with peak wavelength at 600 nm, obtaining a resolution of λ/6 that is well beyond the classical limit.

Fig. 19 shows Spider silk biological superlens and its imaging performance. The used spider silk is from Nephila edulis spider, with diameter about 6.8 um (Fig. 19 a). The spider silk was placed directly on top of the sample surface by using a transparent cellulose-based tape. The gaps between silk and sample was filled with IPA which improves imaging contrast. The silk lens collects the underlying near-field object information and projects a magnified virtual image into a conventional objective lens NA:0.9, Fig. 19b). The spider silk magnifies objects about 2.1x times (Fig. 19c), allowing subdiffraction object to be seen. Blu-ray disc with 100 nm features were clearly resolved by the spider silk (Fig. 19d), under an angular beam incident at around 30 degree. Simulated intensity field in Fig. 19g reveals how the samples was illuminated by the angular jet. In Fig. 19(h), it was also shown that the spider silk produced similar quality of super-resolution image as BTG microspheres, despite the image was slightly rotated by approximately four degree due to fibre direction is not parallel to grating direction (like case in d). This research provides a solid foundation for the development of bio-superlens technology, and it is the first expose of spider silk in this context. It is expected more biological superlenses to be discovered. Indeed, interesting and important recent work by other researchers is heading in that direction. For example, Schuergers et al. reported that spherically shaped cyanobacteria Synechocystis cells can focus light like a micro-lens, which contributes to the cell's ability to sense the direction of lighting source. These cells might work also as a superlens suitable for super-resolution imaging. Very recently, it was reported that live cyanobacteria can form an aggregated metamaterial-style biological lens which focuses light; however, its imaging performance hasn't been evaluated yet which worth being studied in next-step.

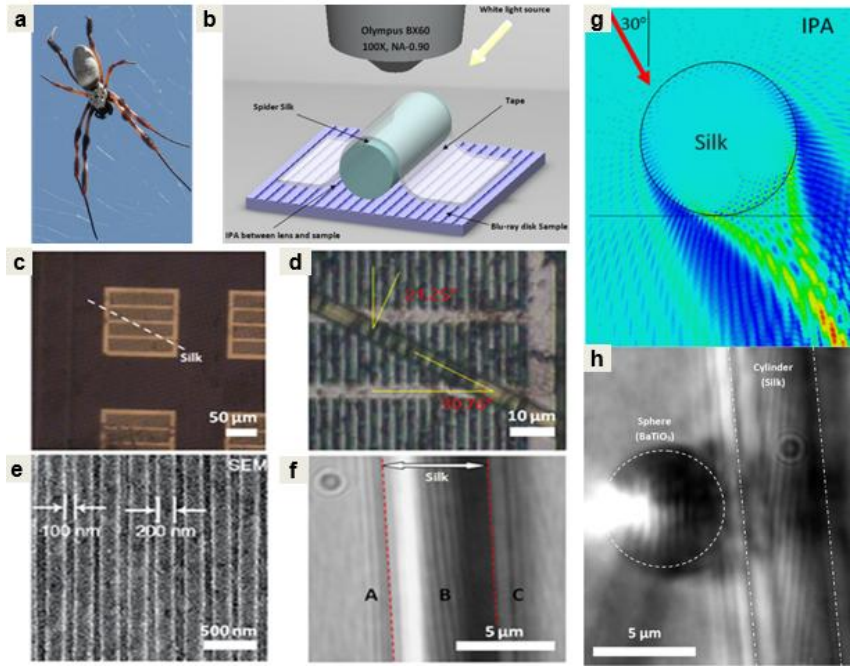

Fig. 19 Spider silk biological superlens and super-resolution imaging. (a) Ne-
phila edulis spider, (b) schematic drawing imaging setup. ( c) Silk on test pattern,
in an angle. (d) Imaging by silk showing magnification and image rotatin effect
(e ) SEM image of Blu-ray disc with 100 nm features, (f) 100 nm features were
resolved, magnified 2.1x by the spider silk cylinder lens. (g) simulated optical
near- field distribution around the silk lens, with beam incident at an angle of 30-
degree (h) A BTG microsphere positioned beside the minor ampullate spider silk
for imaging comparision experiments.

## 6. Devices with integrated microsphere superlens

Due to tiny size of microspheres, it has advantage of being integrated in other microsys-
tems, such as microfluidics and optical fiber system, to form a multifunctional on-chip de-
vice. For example, Yang et al. demonstrated a microfluidic device with an integrated mi-
crosphere-lens-array, as shown in Fig. 20(a-c)[98]. The microspheres are patterned in a
microwell array template, acting as lenses focusing the light originating from a microscope
objective into photonic nanojets that expose the medium within a microfluidic channel.
When a NP is randomly transported through a nanojet, its backscattered light (for a bare Au
NP) or its fluorescent emission is instantaneously detected by video microscopy. Au NPs
down to 50 nm in size, as well as fluorescent NPs down to 20 nm in size, are observed by
using a low magnification/low numerical aperture microscope objective in bright-field or fluorescence mode, respectively. Compared to the NPs present outside of the photonic nanojets, the light scattering or fluorescence intensity of the NPs in the nanojets is typically enhanced by up to a factor of ~40. The experimental intensity is found to be proportional to the area occupied by the NP in the nanojet. The technique is also used for immunodetection of biomolecules immobilized on Au NPs in buffer and, in future, it may develop into a versatile tool to detect nanometric objects of environmental or biological importance, such as NPs, viruses, or other biological agents. In another work by Li and co-workers, microsphere lenses were attached onto an optical fibre probe; the generated PNJ focus was used for real-time manipulation (trapping and releasing) and detection of 85 nm objects like nanoparticles and biomolecules, as shown in Fig. 20(d,e). A multifunctional Lab-on-Chip (LoC) device capable of delivering real-time trapping, detection and super-resolution imaging functionalities was recently proposed and partially experimentally demonstrated by Yan et al., where a coverslip-style microsphere superlens was bonded onto a microfluidic chip on silicon. The trapping function was realized by dielectrophoretic (may also with PNJ effect). When bio-samples flowing through the channel, the objects will be trapped by the dielectrophoretic force at locations beneath the particles, super-resolution images were then obtained through microsphere imaging. The successful realization of proposed device may lead to fundamental changes to biomedical analysis with accuracy in nanometre scales.

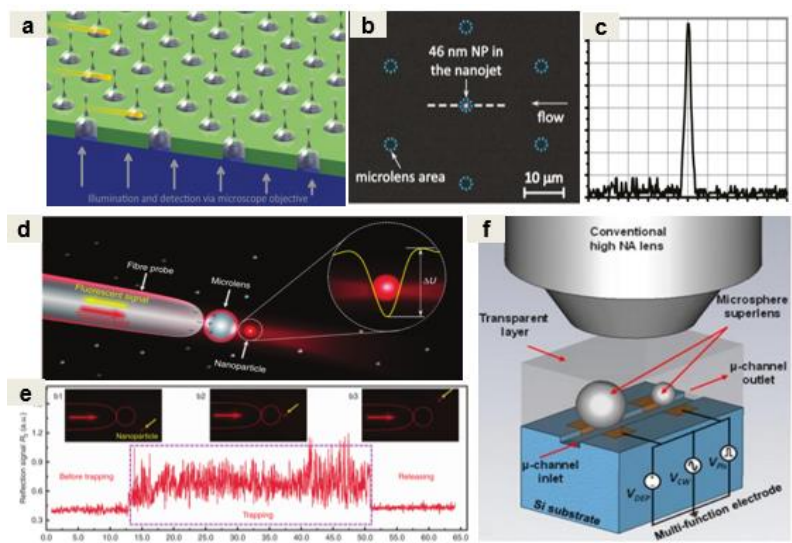

Fig. 20 Device with integrated microspheres. (a) Schematic of microsphere-array-assisted nanoparticle sensing platform. (b) 46 nm fluorescent nanoparticle was detected by microsphere-generated PNJ. (c) corresponding backscattering signal. (d) Schematic illustration of fiber-microsphere-bonded probe detection and trapping of nanoparticles (b) The real-time trace of the reflected signal in the trapping process of an 85-nm fluorescent PS nanoparticle. The insets show the fluorescent images b1 before trapping, b2 during trapping and b3 in the release. (f) A multifunction biochip device with integrated sensing, trapping and super-resolution imaging functionalities.

## 7. Outlook and conclusions

Compared to other super-resolution techniques, the microsphere nanoscopy technique has several distinct features – simple, easy to implement, label-free, high-resolution and compatible with conventional white-lighting imaging. The scanning superlens system has generated possible opportunities for commercialization, a few start-up companies were recently setup using first-generation dielectric superlenses (PS or BTG particles) technology. These systems are likely to suffer from low imaging contrast and quality compared to second-generation particle superlenses based on metamaterials. Development of commercially viable prototype of metamaterial dielectric superlens is of considerable interests in the next-step development of the technique.

For microsphere imaging, one possible way to further improve the system resolution is to use *microfiber* to evanescently illustrate the specimen [99]. Imaging contrast in this case can be greatly improved owing to the limited illumination depth (typical <200 nm) of the evanescent waves; sharp and clear images of nanostructures have been achieved[99]. This idea may worth further exploration for sub-50 nm resolution imag-ing in the future, by extending the evanescent wave illumination approach to micro-spheres; either prism-style or objective-style TIRM (Total Internal Reflection micros-copy) setup can be used [100].

On application side, the breakthrough is likely to come from in-vivo endoscopy application where other super-resolution techniques (STED) cannot be applied. The super-resolution endoscope can provide real-time high-resolution image of internal cells and tissues for the early diagnostics of cancers and other diseases.

On the other hand, studies on new effects generated by particles are emerging and growing, including nanoparticle super-resolution, microsphere super-resonance, micro-sphere array Talbot effect[101], and micro-prism ''photonic hook'[102], etc. The 'photonic hook' effect may be used for nanoscale light switching and guiding in a Photonic Integrated Chip. The factual Talbot effect of a microsphere lens array may be used to develop a far-field particle-lens super-resolution imaging system.

On theoretical side, building a complete theoretical model is still highly desired. Effects such as multi-wavelengths effects and partial and inclined illumination should also be included in the developments. In a long term, we envisage that every microscope user will have the dielectric superlenses in their hand for daily use of microscopes. Besides imaging, the dielectric particle superlens can find applications in nano-focusing, nanolithography, nano-solar energy concentrator, nanochemistry, nanomedicine and more.